\documentclass[twocolumn,preprintnumbers,amsmath,amssymb]{revtex4}
\usepackage{graphicx}
\usepackage{amssymb}

\usepackage{multirow}

\begin{document}

\title{Comment on \textquotedblleft Masking quantum information is
impossible\textquotedblright }
\author{Guang Ping He}
\email{hegp@mail.sysu.edu.cn}
\affiliation{School of Physics, Sun Yat-sen University, Guangzhou 510275, China}

\begin{abstract}
The no-masking theorem (Phys. Rev. Lett. 120, 230501 (2018)) claims that
arbitrary quantum states cannot be masked. Based on this result, the authors
further suggested that qubit commitment is not possible. Here we show that
this connection does not hold in general.
\end{abstract}

\maketitle


Masking the quantum information contained in states $\{\left\vert
a_{k}\right\rangle \in H_{A}\}$, according to Definition 1 of \cite{qbc140},
means mapping them to states $\{\left\vert \Psi _{k}\right\rangle _{AB}\in
H_{A}\otimes H_{B}\}$ such that all the marginal states of $\left\vert \Psi
_{k}\right\rangle _{AB}$\ have no information about $k$. While masking
classical information is possible, as shown in \cite{qbc140}, the same is
not true for arbitrary quantum states, except for some restricted set of
states only.

However, we should note that arbitrary \textit{known} quantum states can
always be masked with the following method. In brief, any known state must
be able to be described by some classical information. Since classical
information can be masked, we can use the scheme for masking classical
information to mask the description of known quantum states.

For example, if the state to be masked is chosen from an alphabet of states $%
\{\left\vert \psi _{1}\right\rangle ,\left\vert \psi _{2}\right\rangle
,...,\left\vert \psi _{i}\right\rangle ,...\}$, we can first convert the
classical index $i$ of the chosen state into a bit-string, then mask it bit
by bit using the scheme in the 3rd paragraph of \cite{qbc140}. That is, the
bit value $0$ is encoded as the entangled state $(1/\sqrt{2})(\left\vert
00\right\rangle +\left\vert 11\right\rangle )$, and $1$ is encoded as $(1/%
\sqrt{2})(\left\vert 00\right\rangle -\left\vert 11\right\rangle )$. Since
none of the subsystems of these entangled states contains the information
about the classical bits, the index of the state being masked is hidden
completely in the correlations, satisfying the definition of masking. If at
a later time the subsystems are brought together and measured collectively
in the Bell basis, then we can learn the index successfully and recreate the
state.

More general, the state of any qubit to be masked can be expressed as%
\begin{equation}
\left\vert \psi \right\rangle =\cos \alpha \left\vert 0\right\rangle
+e^{i\theta }\sin \alpha \left\vert 1\right\rangle .  \label{qubit}
\end{equation}%
Therefore, even if there is no given alphabet of states to choose from, the
quantum information in the qubit is completely characterized by the
classical parameters $\alpha $ and $\theta $. By converting them into
bit-strings, we can also mask them using the above scheme for masking
classical information. There could be doubt that if $\alpha $ and $\theta $
are irrational numbers such as $\pi $ and $\sqrt{2}$, then they cannot be
converted into finite bit-strings so that the above masking scheme cannot be
completed in finite steps. This is not true. The current paper that you are
reading already contains many $\pi $ and $\sqrt{2}$, and it is actually
stored as an electronic file in the computer with a finite number of $0$ and
$1$. That is, by using some kinds of coding (e.g., the ASCII table),
irrational numbers like $\pi $ and $\sqrt{2}$ can be treated as symbols and
represented by finite bit-strings. Alternatively, as long as the description
of the state can be written down on paper, then it can be scanned and stored
as a bitmap image (BMP) file which also consists of a finite number of bits.
The only exception maybe some irrational numbers that cannot be represented
by a symbol or rule, so that some digits of its decimal expansion remain
unknown. But rigorously speaking, a quantum state described
by such numbers is not exactly a \textquotedblleft
known\textquotedblright\ state, so that it falls outside the scope of our current discussion.

It is trivial to show that the same idea also applies to arbitrary known quantum
states not limited to qubits. That is, as long as a quantum state can be described, then it can be mapped into classical bit-strings, and therefore it can be masked.

Some might argue that the above method does not mask the quantum states
directly. Instead, it masks the classical information that describes the
state. Therefore, it may not be considered as the kind of masking studied in
\cite{qbc140}. But it does not matter whether the method can be taken as a
disproof of the no-masking theorem or not. What is important is that, as
shown below, the existence of this method is sufficient to dispute the claim
that no-masking implies no qubit commitment.

As defined in page 4 of \cite{qbc140}, qubit commitment means that Alice
commits to a qubit from a certain set, and later unveils to Bob that she has
indeed committed to that qubit. Here we are not going to discuss the
validity of the claim that \textquotedblleft qubit commitment is not
possible\textquotedblright . What we doubt is the claim of \cite{qbc140}
that \textquotedblleft the no-masking theorem implies that qubit commitment
is not possible\textquotedblright . To this end, we should be aware that the
definition of qubit commitment actually implies that the quantum state being
committed is always known to at least one of the parties. Notably, in the
opening phase, Alice is required to unveil her committed state for Bob to
verify. It is obviously that if none of Alice and Bob knows the description
of the state, then the unveiling and verification cannot be done. Even if
the state is chosen from a set $\{\left\vert \psi _{1}\right\rangle
,\left\vert \psi _{2}\right\rangle ,...,\left\vert \psi _{i}\right\rangle
,...\}$ where the parameters $\alpha $ and $\theta $ in Eq. (\ref{qubit}) of
every $\left\vert \psi _{i}\right\rangle $\ are unknown to Alice and Bob, at
least the index $i$ labelling the committed state should be known.
Therefore, committing to a qubit actually means committing to the classical
description of the state, and qubit commitment can be reduced to bit-string
commitment \cite{qi161,qi197}. As we shown above, masking the classical
description of arbitrary known quantum states is always possible. The
no-masking result does not apply here. Thus, even if qubit commitment is
indeed impossible, the corresponding proof should be standalone or find its
base elsewhere, instead of being considered as a conclusion derived from the
no-masking theorem.

In short, arbitrary known quantum states can be masked by masking the
classical description of the states. Since the states in qubit commitment is
always known, the no-masking theorem does not apply to this scenario.
Therefore, no-masking cannot lead to no qubit commitment.

This work was supported in part by Guangdong Basic and Applied Basic
Research Foundation under Grant No. 2019A1515011048.

\end{document}